\documentclass{cimento}
\usepackage{amsmath}
\usepackage{amsthm}
\usepackage{amssymb}
\usepackage[italian]{babel}
\usepackage[latin1]{inputenc}
\usepackage{epigraph}
\usepackage{expdlist}
\usepackage{epsfig}

\title{Performance study of single undoped CsI crystals for the Mu2e experiment}
\author{R.~Donghia\from{ins:lnf} \from{ins:rm3}
on behalf of the Mu2e calorimeter group}
\instlist{\inst{ins:lnf} Laboratori Nazionali di Frascati dell'INFN.
                      \inst{ins:rm3} Università degli Studi Roma Tre, dipartimento di Fisica.     }

\PACSes{
\PACSit{29.40.Mc}{29.40.Vj}}

\begin{document}

\maketitle

\begin{abstract}
The Mu2e experiment at Fermilab aims to measure the charged lepton flavor violating neutrinoless muon to electron conversion. The goal of the experiment is to reach a single event sensitivity of $2.5 \times 10^{-17}$, to set an upper limit on the muon conversion rate at $6.7 \times 10^{-17}$ in a three years run. For this purpose, the Mu2e detector is designed to identify electrons from muon conversion and reduce the background to a negligible level. It consists of a low mass straw tracker and a pure CsI crystal calorimeter. 

In this paper, the performance of undoped CsI single crystal is reported. Crystals from many vendors have been characterized by determining their Light Yield (LY) and Longitudinal Response Uniformity (LRU), when read with a UV extended PMT, and their time resolution when coupled to a Silicon Photomultiplier. 

The crystals show a LY of $\sim$ 100 photoelectrons per MeV when wrapped with Tyvek and coupled to the PMT without optical grease. The LRU is well represented by a linear slope that is on average 0.6\%/cm. Both measurements have been performed using a $^{22}$Na source.

 The timing performance have been evaluated exploiting minimum ionizing particles, with MPPC readout. A timing resolution better than 400 ps is achieved (at $\sim$~20~MeV).

\end{abstract}
\section{Introduction}
The Mu2e experiment \cite{ref:tdr} at Fermi National Accelerator Laboratory,  will search for the charged lepton flavor violating (CLFV) process of muon conversion in an aluminum nucleus field, $\mu + N (Z, A) \rightarrow e + N (Z, A)$. 

Mu2e measure the ratio, $R_{\mu e}$, between the muon conversion and the muon capture rates, by Al nucleus, with a single event sensitivity (SES) of $2.5 \times 10^{-17}$. This corresponds to set an upper limit of:
\begin{equation*}
R_{\mu e} = \frac{\mu^- \thinspace N(Z,A) \rightarrow e^- \thinspace N(Z,A)}{\mu^- \thinspace N(Z,A) \rightarrow\nu_{\mu} \thinspace N(Z-1,A)} < 6 \times 10^{-17},~(@~90\%~CL),
\end{equation*}
improving four orders of magnitude the previous result from SINDRUM II \cite{sindrum}.

The signature of this neutrinoless conversion process is a mono energetic electron, with an energy slightly lower than the muon rest mass ($\sim$ 104.96 MeV). 

The Standard Model predicted rate for this process is $\mathcal{O}(10^{-52})$, so that,  observation of these processes could be a clear evidence for New Physics.

\section{The Mu2e Calorimeter}
The Mu2e detector consists of a low mass straw tracker and a crystal calorimeter. Both are located inside the evacuated warm bore of the Detector Solenoid in a uniform 1 T magnetic field, that is surrounded by a cosmic ray veto.

The detector is designed to identify the $\sim$105 MeV/c electrons from muons conversion, reducing the background to a negligible level. 

The calorimeter, located behind the tracker, has to provide information about energy, timing and position to validate the charged particle reconstructed by the tracker and perform a particle identification. 

To reach the SES required by the experiment and also maximize the acceptance for $\sim 105$ MeV conversion electrons, a crystal calorimeter with an energy resolution of $\mathcal{O}(5\%)$ and a timing resolution better than 500 ps in the energy region around 100 MeV is required. 
Moreover, the Mu2e environment implies the use of solid-state photodetectors immune to the presence of the magnetic field.

The baseline calorimeter is composed by 1400 pure CsI crystals, distributed in two annular disks and readout by two silicon photomultipliers (SiPMs) each.

In this paper the tests done on the light yield (LY) and the longitudinal response uniformity (LRU) of single CsI crystals are reported.  Moreover, the timing performance has also been evaluated for some crystals, by coupling them to a SiPM.

\subsection{Pure Csi crystals}
Undoped Cesium Iodide (CsI) is a slightly hygroscopic crystal with an emission maximum at 315 nm, characterized by a relatively short decay time of $\sim$~20~-~30~ns \cite{csi}. Together to this fast component, a much slower component with a decay time of about 1~ms is present which represents less than 15\% of the total light output. The intensity of this slow component depends on the purity of the crystal since contamination with certain elements tends to degrade the fast-to-total ratio. This contribution is practically negligible in our CsI samples.

\subsection{Silicon Photomultiplier, MPPC}
Due to the high magnetic field, the CsI crystals readout has to be made by high-gain solid-state photodetectors, such SiPMs.

SiPMs are photon-counting devices made by one planar matrix of several avalanche photodiode (APD) pixels of the same shape, dimension and construction features that are operated in Geiger mode, with an inverse polarization above the breakdown. Each pixel is coupled to a quenching resistor \cite{mppc}. 

At the wavelength emission peak of pure CsI (315 nm) the UV extended Hamamatsu SiPM, called Multi-Pixel Photon Counter (MPPC) is a good choice for the Mu2e calorimeter. In particular, the performance of S13361 series MPPCs are under study. 
It allows precision measurements, using the TSV (Through Silicon Via) technology \cite{tsv_mppc}. There is no wire bonding, so the package outline is very close to the MPPC array. The outer gap from active area edge to package edge is only 0.2 mm. 

\section{Tests on single pure CsI crystals coupled to a UV extended PMT}
To test the crystal production quality, we have procured 13 samples of pure CsI crystals from different high quality producers: 2 from Opto Materials (Italy), 7 from ISMA (Ukraine) companies, both with a crystal dimension of $(3\times 3 \times 20)$ cm$^3$ and four additional longer crystals $(2.9\times 2.9 \times 23)$ cm$^3$ from ISMA. The measurements have been performed at the INFN National Laboratory of Frascati (LNF) using a dedicated station for crystals testing. 

\subsection{Experimental setup}
To study the LY and LRU of each crystal, we have used a low intensity collimated $^{22}$Na source which irradiates the crystal in a region of few mm$^2$. The $^{22}$Na source produces 511 keV electron-positron annihilation photons and it is placed between the crystals and a small tagging system, constituted by a ($3\times 3\times 10$)~mm$^3$ LYSO crystal, readout by a ($3\times 3$) mm$^2$ MPPC.

One of the two back-to-back photons produced by the source is tagged by this monitor, while the second one is used to calibrate the crystal under test, which is readout by means of a 2'' UV extended photomultiplier tube (PMT) from ET Enterprises. This PMT has a quantum efficiency of $\sim 30\%$ at 310 nm, which is the wavelength where the undoped CsI reaches the emission maximum. The whole system is inside a light tight black box.

The data acquisition system is composed by a trigger board, which starts recording events applying a threshold of 20 mV on the tag signal, and a CAEN DT5751 digitizer at 10$^9$ samples per second, which acquires both tag and test signals.

For each crystal, a longitudinal scan is done irradiating eight points, of 2~cm step from the readout system. In the scan, the source and the tag are moved together along the axis of the crystal under test. 
All crystals have been tested when wrapped with a reflector material, which covers both the four surfaces along the longitudinal axis and the side opposite to the readout system. 

\begin{figure}[b!]
\begin{center}
\epsfig{file=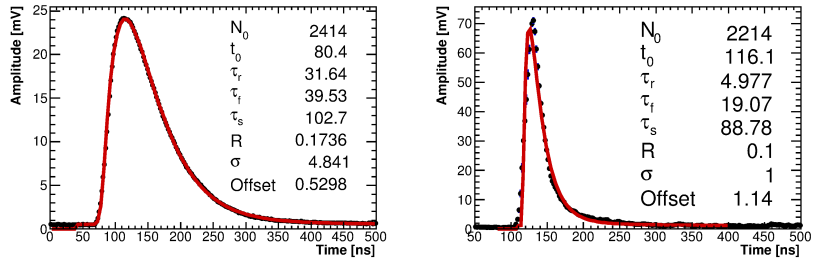,width=.8\textwidth}
\caption{\emph{Digitized waveform for few events of the tag (top left) and of one Tyvek wrapped Opto Materials 01 crystal (top right) acquired by the CAEN digitizer.}}
\label{Fig:pulse}
\end{center}
\end{figure}

The digitizer has 1024 samples in the acquisition window, each sample corresponding to 1 ns. Examples of the pulse shapes, obtained for both tag and CsI crystals, are shown in Figure \ref{Fig:pulse}.  The generic emission time distribution for a scintillator can be described as a fast component, generated by a two-step scintillation mechanism (absorption, emission) and a slow component \cite{func}:
\begin{equation}
E(t)=\frac{\frac{e^{-t/\tau_f}-e^{-t/\tau_r}}{\tau_f-\tau_r}+\frac{R}{\tau_s}e^{-t/\tau_s}}{1+R} ,
\end{equation}
where $\tau_f$, $\tau_s$, $\tau_r$ are the time constants of the fast and slow scintillation process and of the rising part, respectively. $R$ is the ratio between the slow and the fast component. If we assume that the time resolution of our system can be described by a Gaussian, then the resulting distribution used to fit the waveforms is the convolution of $E(t)$ with a Gaussian, as follows:
\begin{equation}
V(t) = \frac{1}{1+R}\left[\frac{\tau_f f(t,\tau_f)-\tau_r f(t,\tau_r)}{\tau_f-\tau_r}+R f(t,\tau_s)\right] \thinspace,
\label{eq:func}
\end{equation}
where:
\begin{equation}
f(t,\tau)=\frac{1}{2\tau}\left[1+\mathrm{erf}\left(\frac{1}{\sqrt{2}}\left(\frac{t}{\sigma}-\frac{\sigma}{\tau}\right)\right)\right]e^{-(t/\tau-\sigma^2/2\tau^2)}
\end{equation}
$erf$ is the error function, defined as:
\begin{equation}
erf(x)=\frac{2}{\sqrt{\pi}}\int_0^x e^{-t^2} \, dt
\end{equation}
and $\sigma$ is the Gaussian standard deviation.

The profiles of the waveforms have been fit with Eq.~\ref{eq:func} to evaluate the decay time of each crystal. Fit results are reported in Figure~\ref{Fig:pulse}. Since CsI has a very small slow component, $R$ parameter has been fixed to 0.1, while the resolution function of our system has been set to $\sigma = 1$ ns.
As shown in the same figure, in our setup, signals produced in CsI crystals are typically within 300 ns from the trigger, with a 50 ns delay offset, so that the charge $Q$ is obtained integrating in the range ($50\div 300$)~ns. The baseline is evaluated using the interval region above 700 ns. 
\begin{figure}[!h]
\begin{center}
\epsfig{file=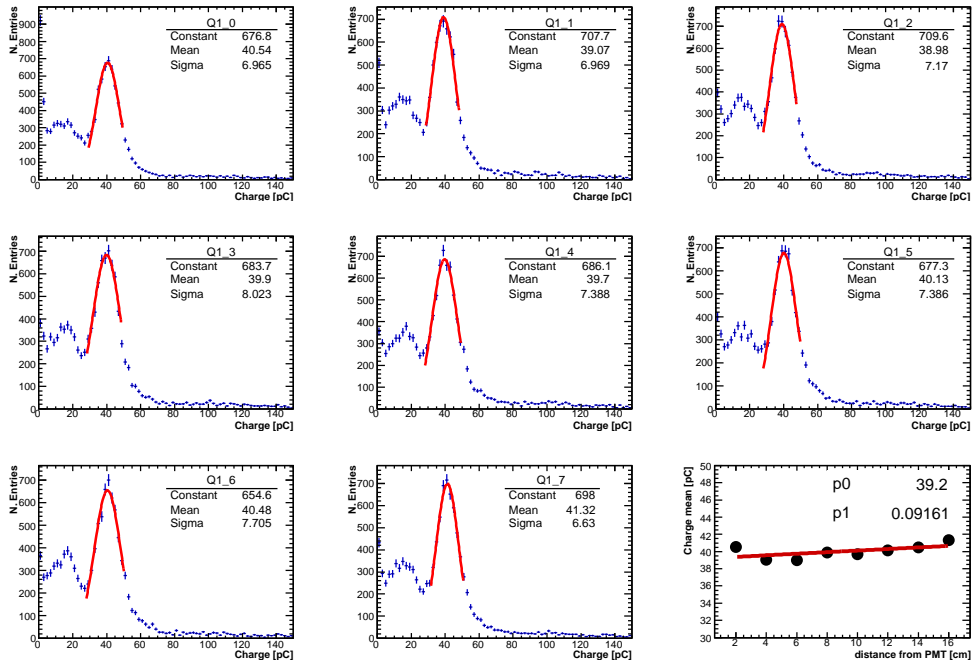,width=1\textwidth}
\end{center}
\caption{\emph{Fits used to extract the charge peak position during the scan of the Opto Materials 01 crystal, wrapped with Tyvek and plot of these values as function of distance.}}
\label{Fig:charge1}
\end{figure}

In Figure~\ref{Fig:charge1}, the charge distributions for one of the crystals under test (Opto Materials 01), wrapped with Tyvek, are reported for the eight/one points of a single scan. The charge spectra are very clean and the peak due to the 511 keV photon is clearly visible. 
A Gaussian fit is applied to extract the mean values ($\mu_{Q1}$): these values are then reported as a function of the distance of the source from the PMT, obtaining a linear slope parametrized as $39\thinspace\mathrm{pC} + 0.09\thinspace\mathrm{pC/cm}$ (Fig.~\ref{Fig:charge1}, bottom right). 

\subsection{Light emission and longitudinal response uniformity performance}
All crystals have been tested with the $^{22}$Na using just one orientation with respect to the readout system. For some of the crystals, the effect of the optical grease contact with the PMT has also been studied.

A reflector material wrapping is needed to improve the detection efficiency of scintillation photons. The Opto Materials pure CsI crystal number 02 has been tested with different wrapping materials: aluminum (Al), Tyvek, Teflon. The charge distributions have been fit with a gaussian function to extract the peak position and evaluate the LY, defined as the number of photoelectrons produced per MeV, $N_{p.e.}/MeV$:
\begin{equation}
\frac{Np.e.}{MeV} = \frac{\mu_{Q1 [pC]}}{G_{PMT}\cdot E_{\gamma} [MeV] \cdot q_{e^-} [pC]},
\label{eq:ly}
\end{equation}
where $q_{e^-}=1.6\times10^{-19}$ pC is the charge of the electron, $E_\gamma = 511$ keV is the energy of the annihilation photon and $G_{PMT}=3.8\times 10^6$ is the PMT gain.

The Al wrapping provided the worst LY ($\sim$79 N$_{p.e.}$/MeV with the source in the central position), while the best performance, for every scan point, has been obtained with Teflon ($\sim$91 N$_{p.e.}$/MeV) and Tyvek wrapping ($\sim$~89 N$_{p.e.}$/MeV), which provide a LY increase of a factor about 20\% with respect to the configuration with Al. Therefore, tests on all the other crystals have been carried out with Tyvek wrapping, due to the fragility and difficulty to repair Teflon, especially when in presence of optical grease \cite{biodola}. 

\begin{figure}[!t]
\begin{center}
\epsfig{file=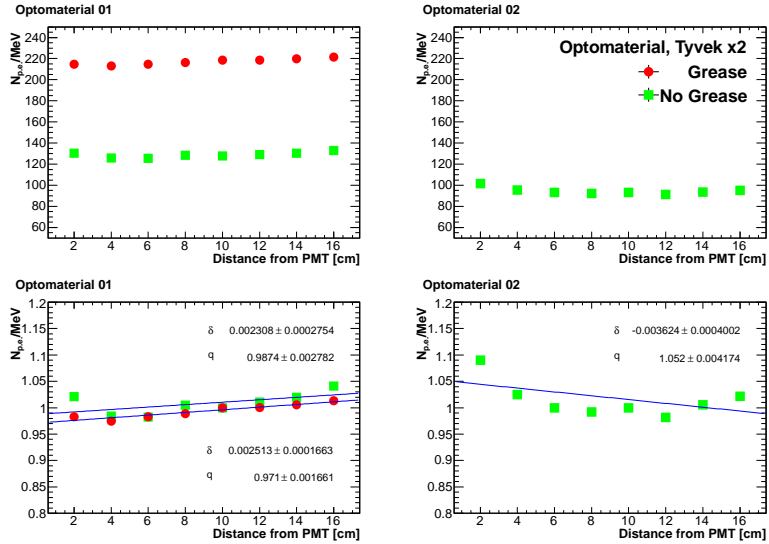,width=.8\textwidth}
\end{center}
\caption{\emph{Top: Number of photoelectrons per MeV produced in the Opto Materials 01 (left) and 02 (right) crystal samples as function of the distance of the source from the PMT. Bottom: LY normalized to the number of photoelectrons per MeV produced in the central scan position and fitted with a line function to evaluate the LRU ($\delta$) in \%/cm.}}
\label{Fig:confr_sardi}
\end{figure}

Examples of scan results are reported in Figure \ref{Fig:confr_sardi}. In the top left figure, a comparison between optical grease and air coupling is also shown, an improvement of about 80\% is clearly visible for all the scan points.

\begin{figure}[!htbp]
\begin{center}
\epsfig{file=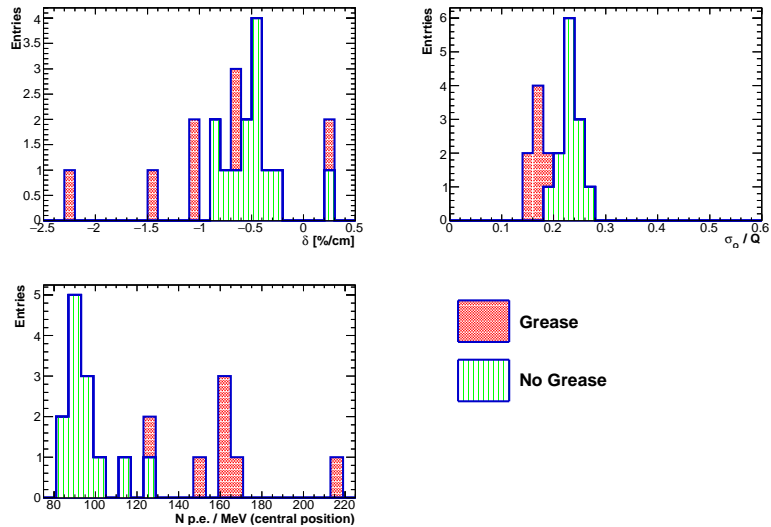,width=0.8\textwidth}
\end{center}
\caption{\emph{Slopes distribution (in \%/cm) provided by the linear fit on the LY normalized as function of the distance from the PMT (top left), resolution distribution (top right) and LY distribution  (bottom left) obtained with the source in the central scan position (at 10 cm far from PMT) for all the crystal samples tested.}}
\label{Fig:tefl38}
\end{figure}

To summarize, tests on all the 13 crystals show that:
\begin{itemize}
\item relevant differences between crystals from the same company exist. For instance, we observe a 45\% better LY with Opto Materials sample 01 with respect to sample number 02 (Fig.~\ref{Fig:confr_sardi});
\item similar performance for ISMA crystals, both long and short ones, and Opto Materials 02, while Opto Materials 01 has much better uniformity and LY (130~N$_{p.e.}$/MeV with respect to $\sim$ 100~N$_{p.e.}$/MeV);
\item larger signals are observed closer to the PMT, because of the collection of direct light;
\item the charge resolution is $\sim$18\% ($\sim$25\%) with (without) optical grease (Fig.~\ref{Fig:tefl38}).
\end{itemize}
To resume features of all the 13 crystals tested, the LY obtained with the source in the central scan position has been reported in Figure~\ref{Fig:tefl38} (bottom left). To evaluate crystals LRU, the LY, normalized to its value in the central position, has been plotted as a function of the scan position and fit with a linear function (Fig.~\ref{Fig:confr_sardi}). 
Fits angular coefficients are reported in Figure~\ref{Fig:tefl38}, showing a LRU better than 0.5\%/cm in most of the crystals.

\section{Single channel Timing performance}
A $(3\times 3 \times 20)$ cm$^3$ sample from each company, Opto Materials and ISMA, have been coupled to a SPL MPPC and tested with cosmic rays to evaluate the timing resolution. 

\subsection{Experimental setup}
In order to optimize the light collection, crystals have been wrapped with 100 $\mu$m-thick Tyvek foils, covering both the four faces along the crystal axis and the side opposite to the readout system. Each crystal has then placed between two small plastic scintillators, \emph{fingers}, perpendicular to each other and positioned one below and one above the crystal under test. In this way, the two finger coincidence covers 1 cm$^2$ area on the long surface of the crystal. For each crystal, the effect of the Rhodosil Paste 7 optical grease coupling has also been studied. The whole system has been assembled inside a light tight black box.

The data acquisition system is composed by a trigger board, that makes the coincidence between the two discriminated finger signals, and a CAEN DT5751 digitizer at 1~Giga samples per second, which acquires finger and crystal signals.

The goal of this test is to measure the time resolution achieved at the energy released by a minimum ionizing particle (MIP) in the crystal. In order to set the energy scale, we compare the charge spectra of the MIP with that from a radioactive source. Our $^{22}$Na source emits 511 keV back-to-back photons from annihilation and its charge spectrum has a corresponding mean value of ($11.57 \pm 0.17$) pC. For the cosmic rays charge plot, the most probable value is ($463.3 \pm 1.13$) pC. Comparing these two values, the energy released by a MIP in a crystal results to be around 20 MeV.

The amplifier used for the CR test was a prototype version with a gain of 3.


\subsection{Timing analysis}
\begin{figure}[t!]
\centering
\includegraphics[width=9cm]{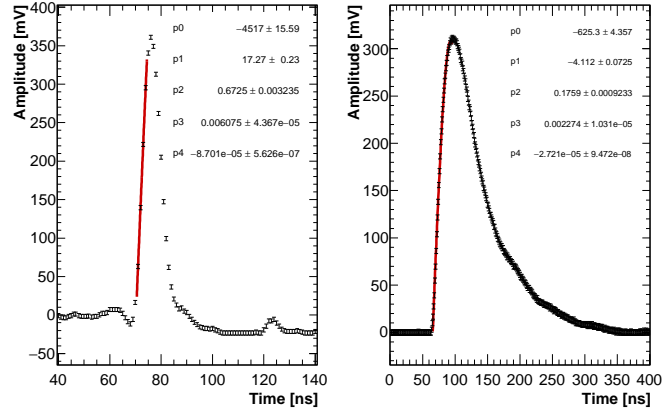}
\caption{\emph{Signals of the top finger (left) and of the Opto Materials 01 crystal (right) acquired by the CAEN DT5751 digitizer at 1 Gsps rate. Also the 4th order polynomial fit is shown.}}
\label{fig:signal}
\end{figure}

Waveform examples obtained for one finger and for the crystal, are shown in Figure~\ref{fig:signal} (left and right respectively). 
To extract the time, the maximum value of pulse height for the finger and crystal signals has been evaluated and then a fit with a 4$^{\mathrm{th}}$ order polynomial function has been performed between the times at two fixed thresholds: $i)$ the position of the time sample corresponding to a pulse height of 10 mV above the signal baseline and $ii)$ the one at maximum pulse height less 1 ns ($t_{max}$ - 1). The fit is shown in Figure~\ref{fig:signal}. 

Both for fingers and crystals the measured time is taken at a constant fraction CF, set at 25\%, of the maximum signal amplitude. In order to eliminate the jitter due to the trigger, the half sum of the finger time has been subtracted:
\begin{equation}
t= t_c - \frac{t_1+t_2}{2}~,
\end{equation}
where $t_1$ and $t_2$ are the time of the bottom and top fingers respectively and $t_c$ is the crystal time. The time jitter of the trigger, $(t_1+t_2)/2$, is evaluated as the $\sigma_f$ provided by the gaussian fit that is $\sim (168 \pm 6)$ ps.

\begin{figure}[h!]
\centering
\includegraphics[width=8cm]{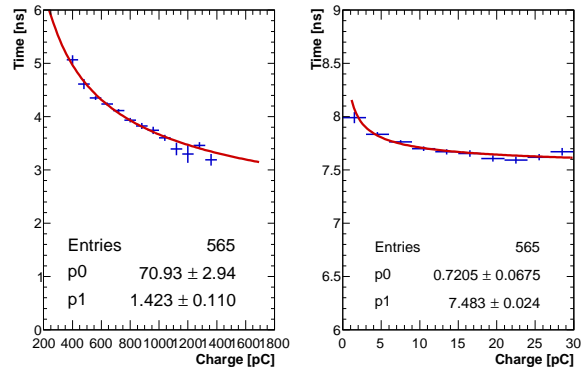}
\caption{\emph{Left: crystal time evaluated as $t = t_{c} - \frac{t_1 + t_2}{2}$ as a function of the signal charge. Where $t_1$ and $t_2$ are the bottom and top finger times respectively. Right: half difference of finger times as a function of the signals charge. The fit reported is used to evaluate the slewing correction.}}
\label{fig:sl_corr}
\end{figure}

The detector timing properties are determined primarily by time slewing (or time walk) resulting from the signal rise time, shape and amplitude. The dependence of the time, $t$, from the charge, $Q$, is shown in Figure~\ref{fig:sl_corr}. This behavior is described by the function: 
\begin{equation}
t = \frac{p0}{\sqrt{Q}}+p1 \thinspace
\label{eq:sl}
\end{equation}
where $p0$ and $p1$ are parameters evaluated by a fit to the dependence in FIg.~\ref{fig:sl_corr} minimizing the $\chi^2$. 

\begin{figure}[h!]
\centering
\begin{tabular}{cc}
\includegraphics[width=5.5cm]{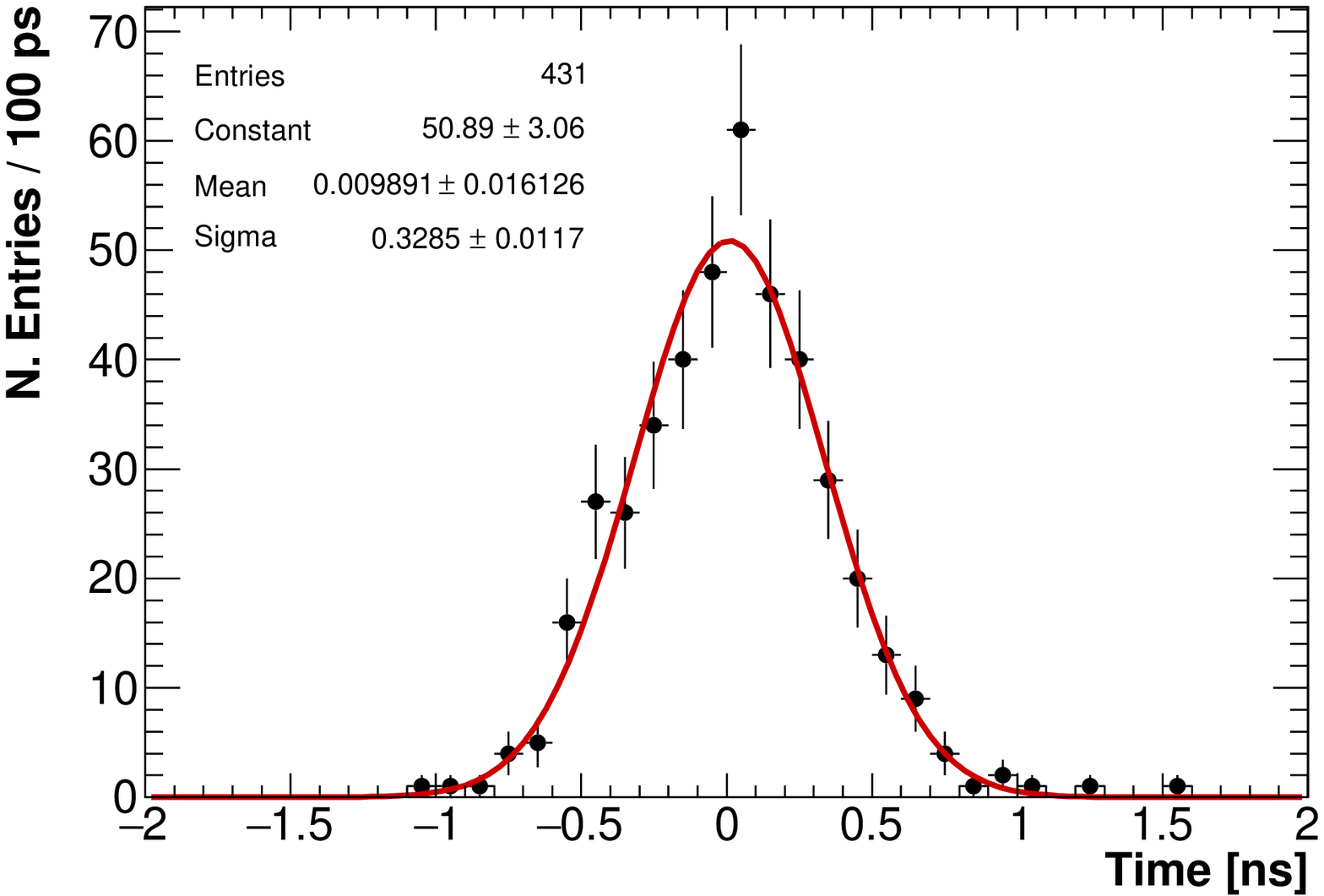}&
\includegraphics[width=5.5cm]{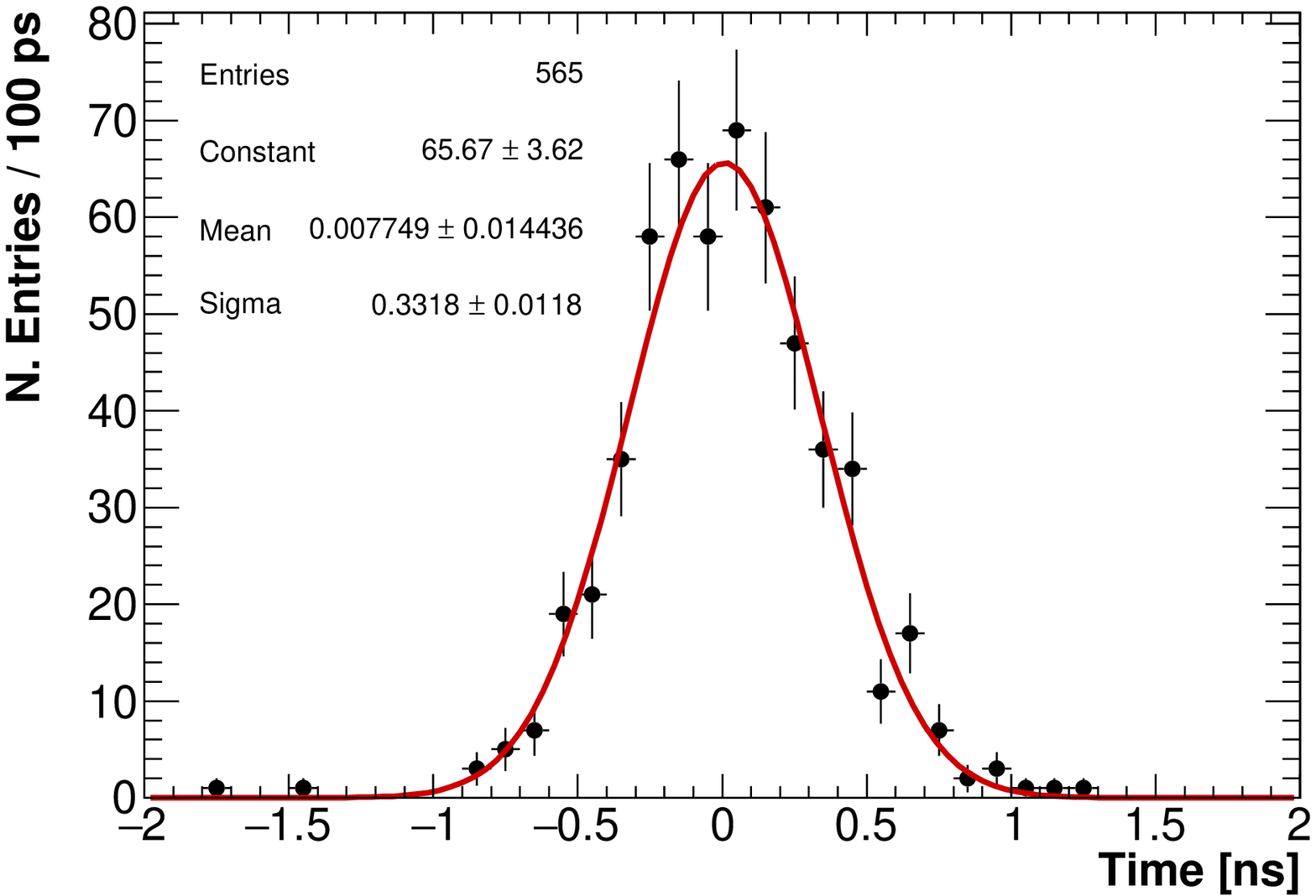}\\
\end{tabular}
\caption{\emph{Time distributions obtained with the constant fraction method after the slewing correction of the Opto Material 01 crystal, coupled to the SPL MPPC with optical grease. This crystal has been tested wrapped with Teflon (left) or Tyvek (right).}}
\label{fig:res_spl_sipm}
\end{figure}

After the time slewing correction, the time distributions of the Opto Material 01 crystal tested, wrapped with Tyvek and Teflon and coupled with and without optical grease to the SPL MPPC, are reported in Figure~\ref{fig:res_spl_sipm}. The time resolution, $\sigma_c$, is the "Sigma" value of the gaussian fit, reported in the same figure.

The best performance is obtained with optical grease coupling: ($328 \pm 12$) ps with Teflon and ($333 \pm 12$) ps with Tyvek wrapping. Coupling in air deteriorates the resolution: ($409 \pm 16$) ps with Teflon and ($455 \pm 13$) ps with Tyvek wrapping, as expexted by the loss of collected light. Then the time resolution after jitter subtraction is evaluated as: 
\begin{equation}
\sigma = \sqrt{(\sigma_c^2 - \sigma_f^2)}
\end{equation}

In Table~\ref{tab:time_res}, all the jitter subtracted time resolutions obtained testing crystal + SPL MPPC exploiting cosmic rays are summarized. 
\begin{table}[b!]
   \begin{tabular}{ccccc}
    \hline
    	& Tyvek 		& Tyvek and grease  & Teflon 	       & Teflon and grease\\ 
    \hline
 Opto Materials 01    	& $\sim$ 410    & $\sim$ 270 	        & $\sim$ 375   & $\sim$ 260 \\
 Opto Materials 02    	&	     -		& 	$\sim$ 280	&	-	       & - \\
 ISMA 05          		&	     -		& $\sim$ 265	 	&      - 	       &  -\\
    \hline
  \end{tabular}
 \caption{\emph{Time resolution, at $\sim$ 22 MeV, given in ps. The value reported are jitter subtracted. The symbol "-" means that this case has not been measured.}}
  \label{tab:time_res}
\end{table}

Another kind of Hamamatsu MPPCs with a different cover layer have also been tested. These are the so called Micro Film MPPCs.

\begin{figure}[h!]
\centering
\includegraphics[width=7.5cm]{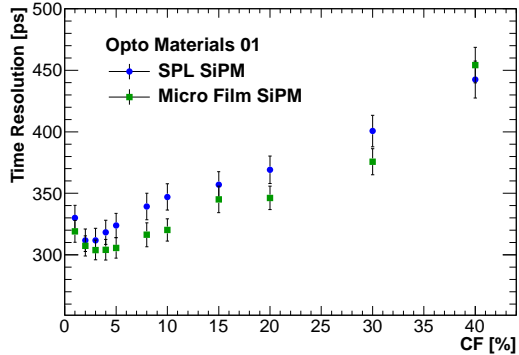}
\caption{\emph{Time resolution as a function of the constant fraction threshold value of the Opto Material 01 with grease coupling. For both the SiPMs, the best resolution is obtained with CF~=~3\%.}}
\label{fig:cf}
\end{figure}

In this final test, the time has been evaluated in a similar manner, by using two variable thresholds for the fit range (the fit function remains a 4th order polynomial) at respectively 0.1\% and 85\% of the signal maximum amplitude. These threshold values are obtained performing a scan and minimizing the time resolution.

Therefore, a scan on the time resolution as a function of the CF threshold has been carried out (Fig.~\ref{fig:cf}). The CF value used for all tests has been chosen as the value which minimizes the time resolution of the crystal, that is CF~=~3\%. For this purpose, only the Opto Material 01 coupled with grease to both SPL and Micro FIlm photosensors has been tested,  the optimized CF value is similar in both cases. 

For the same configuration the SPL MPPC (the Micro Film SiPM) shows an improvement of $\sim$ 2\% ($\sim$ 8.5\%) of the time resolution with respect to the previous method.

\section{Conclusion}
All tested undoped CsI crystals show a good LY $\sim(100 - 130)$ N$_{pe}$/MeV, increased by a factor $\sim$~1.7 when coupled with grease. The LRU has an average of $\sim$~0.6\%/cm. 

The time resolution obtained with cosmic rays is $\sim$~375 ps ($\sim$270 ps), without (with) grease at $\sim$~22 MeV (energy deposited by a minimum ionizing particle in a pure CsI crystal) using Hamamatsu MPPC as readout. 

\acknowledgments
This work was supported by the EU Horizon 2020 Research and Innovation Programme under the Marie Sklodowska-Curie Grant Agreement No. 690835.

\end{document}